\documentclass[review]{elsarticle}
\usepackage[utf8]{inputenc}
\usepackage{amsmath, amsfonts}
\usepackage{times}
\DeclareUnicodeCharacter{0301}{\'{e}}
\usepackage{natbib}
\usepackage{epsfig}
\graphicspath{{./images/}}
\usepackage{geometry}
\usepackage{adjustbox}
\usepackage{multirow}
\usepackage{multicol}
\usepackage{longtable}
\usepackage{hyperref}
\usepackage{marvosym}
\usepackage{placeins}
\usepackage{pbox}
\usepackage{float}
\date{} 
\usepackage{tabu}
\usepackage{subcaption}
\usepackage{url}
\usepackage{booktabs}
\usepackage{graphicx} 
\usepackage{array}    
\usepackage[flushleft]{threeparttable}
\usepackage{amssymb}
\usepackage[none]{hyphenat}
\usepackage{setspace}
\makeatletter
\def\ps@pprintTitle{%
 \let\@oddhead\@empty
 \def\@oddfoot{\reset@font\hfil\thepage\hfil}
 \let\@evenhead\@empty
 \let\@evenfoot\@oddfoot}
\makeatother

\begin{document}

\begin{frontmatter}

\title{Expanding the Katz Index for Link Prediction: A Case Study on a Live Fish Movement Network}

\author[label1]{Michael-Sam Vidza}
\ead{mvidza@bournemouth.ac.uk}
\address[label1]{Bournemouth University, Computing \& Informatics, Fern Barrow, Poole, Bournemouth, BH12 5BB, Dorset, United Kingdom}
\author[label1]{Marcin Budka} \ead{mbudka@bournemouth.ac.uk}
\author[label1]{Wei Koong Chai} \ead{wchai@bournemouth.ac.uk}

\author[label2]{Mark Thrush} \ead{mark.thrush@cefas.gov.uk}
\address[label2]{Centre for Environment, Fisheries and Aquaculture Science (Cefas), Barrack Road, Weymouth, DT4 8UB, , Dorset, United Kingdom}
\author[label2]{Mickael Teixeira Alves} \ead{mickael.teixeiraalves@cefas.gov.uk}

\cortext[cor1]{Corresponding author}

\begin{abstract}
In aquaculture, disease spread models often neglect the dynamic interactions between farms, hindering accuracy. This study enhances the Katz index (KI) to incorporate spatial and temporal patterns of fish movement, improving the prediction of farms susceptible to disease via live fish transfers. We modified the Katz index to create models like the Weighted Katz Index (WKI), Edge Weighted Katz Index (EWKI), and combined models (e.g., KIEWKI). These incorporate spatial distances and temporal movement patterns for a comprehensive aquaculture network connection prediction framework. Model performance was evaluated using precision, recall, F1-scores, AUPR, and AUROC. The EWKI model significantly outperformed the traditional KI and other variations. It achieved high precision (0.988), recall (0.712), F1-score (0.827), and AUPR (0.970). Combined models (KIEWKI, WKIEWKI) approached, but couldn't surpass, EWKI performance. This study highlights the value of extending Katz index models to improve disease spread predictions in aquaculture networks.  The EWKI model's performance demonstrates an innovative and flexible approach to tackling spatial challenges within network analysis.
\end{abstract}


\end{frontmatter}

\noindent

\section{Introduction}
Networks are a common approach used to represent interconnected systems such as as biological \cite{jeong2001lethality}, social \cite{newman2003structure}, computer \cite{boccaletti2006complex}, transport \cite{lin2013complex}, and climate \cite{donges2009complex}. Networks consist of nodes that represent entities, and links that denote the relations or interactions between these nodes. Network analysis studies these systems by examining their underlying structures and providing insights into their inherent dynamics. Link prediction is a key component in complex network analysis used to estimate the likelihood of a link that will exist between two nodes based on observed links and attributes of the nodes \cite{getoor2005link}. Link prediction has been applied in various fields such as friend recommendation \cite{nandi2013survey, adamic2003friends}, protein-protein interaction (PPI) \cite{franceschini2012string, lei2013novel,szklarczyk2015string}, transport planning \cite{lu2010link}, and e-commerce recommendation \cite{lu2010link, adamic2003friends}. In aquaculture, predicting the likely existence of links between fish farms based on the movement of live fish can contribute to better understand and manage disease spread \cite{kiss2006network}.\\
Aquaculture is recognised as an important contributor to world food production \cite{agriculture2000state} and plays a role in producing fish for restocking and aesthetic use \cite{peeler2011application}. However, disease poses a significant threat to the sustainability of aquaculture \cite{jennings2016aquatic}. Understanding how pathogens spread between sites is critical for implementing effective biosecurity, surveillance, and control measures to minimise the impact of disease. The transmission of pathogens between fish farms through various modes, including transport, river, local, and formite transmission  plays a critical role in the spread of diseases. Transport transmission, such as road haulage, involves moving live fish and potentially pathogen-carrying water or equipment between sites, and has been identified as a major cause of disease spread in European aquaculture \cite{murray2006model, gustafson2007hydrographics}. Moreover, river transmission offers a distinct pathway for disease spread, where pathogens released by infected fish can be carried downstream. This transmission type encompasses multiple mechanisms, including pathogen flow with river currents and carrier movements, making it a significant factor in disease dynamics in aquaculture networks \cite{skall2005viral, taylor2010koi}. In addition to these primary transmission modes, local and fomite transmissions also contribute to the spread of diseases. Local transmission involves movements of staff and shared equipment \cite{brennan2008direct}, while fomite transmission accounts for uncontrollable factors like the movement of predators and scavengers \cite{willumsen1989birds}.\\
While there are several factors contributing to the risk of pathogen transmission, movement of live fish is the most significant \cite{murray2013epidemiology,peeler2004qualitative,taylor2011modelling}. Network analysis has been applied to improve understanding of live fish movement and interaction patterns between fish farms. Studies by Green et al. \cite{green2009small}, Tidbury et al. \cite{tidbury2020comparative} and Murray et al. \cite{jones2019contact} have explained the structure of fish movements and their complex interactions. These studies have paved the way for the development of predictive epidemiological and management models applied to aquaculture networks \cite{jonkers2010epidemics, guilder2023aquaculture}. However, these models rely on static historical connections, which limit their ability to respond to the dynamic nature of fish movements by failing to account for new trading relationships that may result during a disease incursion resulting from disease controls applied by industry regulators. To address this gap, our study extends existing network frameworks by incorporating temporal and dynamic features, aiming to improve disease spread model into a more agile tool that can anticipate and adapt to the changing connections between fish farms.\\
Here, link prediction methods are explored to enhance the predictive capacities of existing aquaculture network models. Specifically, similarity-based models are particularly notable for their intuitive nature and computational efficiency \cite{lu2011link}. There are three primary categories of similarity-based models: local, global, and quasi-local indices, each providing unique insights into network dynamics. Local indices, such as common neighbours \cite{newman2001clustering} and Jaccard's coefficient \cite{chowdhury2010introduction}, primarily focus on the immediate neighbourhood of nodes. While these approaches offer insights based on direct connections between nodes, they may overlook the broader network dynamics and, as a result, limit the understanding of complex interactions in the network. In contrast, global indices like the Katz Index \cite{katz1953new} take into account the entire network structure, including both direct and indirect connections at the cost of increased computational complexity. Quasi-local indices, such as the local random walk \cite{liu2010link}, provide a trade-off between local and global methods. Here, the use of the Katz index provides a comprehensive understanding of the aquaculture network identifying potential fish movement patterns beyond immediate connections.
Although the Katz index has been proven effective in network analysis \cite{liben2003link, lu2011link, gao2015link}, its application in aquaculture (and more widely in terrestrial livestock production) is limited; the approach does not consider the spatial distance between farms. In aquaculture, the proximity of farms significantly influences the likelihood of trade. Farms that are geographically closer are more likely to have frequent movements between them, increasing the potential for disease spread \cite{murray2005framework, austin2012infectious}. Conversely, distant farms may have fewer interactions and pose a lower risk of transmitting diseases. In practice, the management of a disease outbreak can result in movement restrictions or changes in trade practices between nearby farms. In such a context, the conventional Katz Index overlooks geographic distance and relies solely on walks within the network, resulting in inaccurate predictions and ineffective disease control measures.\\
This study acknowledges the dynamic nature of interactions in fish farms and the limitations of the current static models. To address these issues, we developed an innovative approach that extends the Katz Index by incorporating spatial weighting factors based on the geographical distances between fish farms. Furthermore, we incorporated temporal dynamics into the models, enabling predictions of future connections based on historical data and recent trends. Results demonstrate that the novel approach substantially improves the prediction of connections in aquaculture, providing a dynamic, responsive, and widely applicable tool for disease management and network analysis.

\section{Material and methods}
\subsection{Experiment setup}
\begin{figure}[H]
\centering
\includegraphics[width=0.85
    \textwidth]{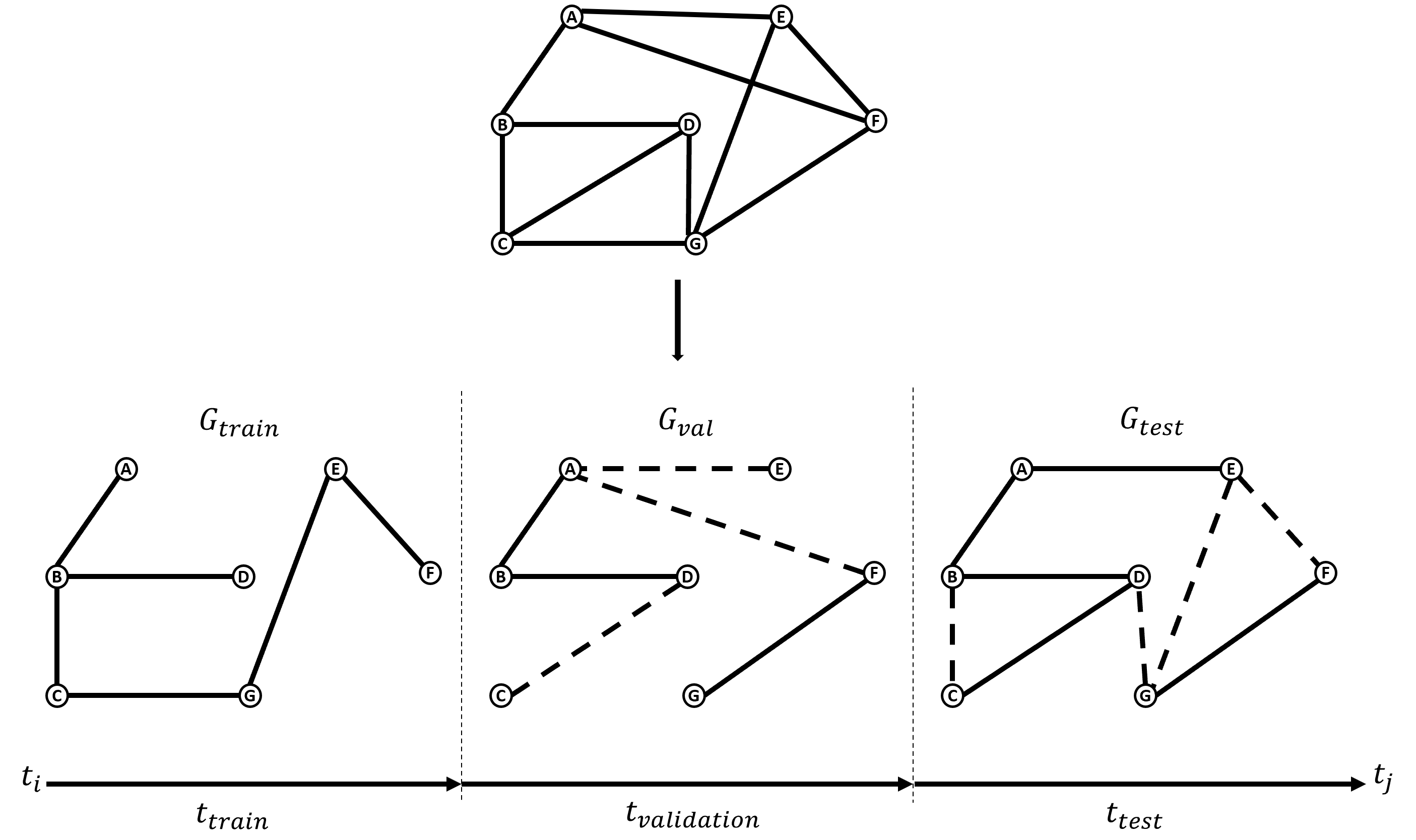}
\caption{Temporal network graph of live fish movements over time. The nodes represent individual fish farms. Solid lines indicate observed movements of fish between farms within the time interval ($t_i$ to $t_j$) and dotted lines denote the predicted future movements for the interval. The horizontal axis represents the timestamps, showcasing the evolution of the network: training ($G_{train}$), validation ($G_{val}$), and test ($G_{test}$) set.}
\label{fig:graph}
\end{figure}

The live fish movement network is represented as $G=(V,E)$ where $V$ represents a set of nodes, corresponding to fish farms, while $E$ represents the set of edges corresponding to the movement between farms. Each edge $(u,v,t) \in E$ represents a connection between nodes $u$ and $v$ at a specific timestamp $t$, showing the movement of fish between farms at different times.  The temporal feature of edges reflects the dynamic nature of fish farm interactions and movements over the period $t_i$ and $t_j$, where $t_i$ and $t_j$ are specific time points with $t_i \leq t_j$. To streamline the network analysis, self-loops were removed. Self-loops are edges that connect a node to itself, illustrating movements of live fish from a farm back to the same farm. Since these movements do not contribute to new connections between different farms, they were removed from the dataset. The network was then divided into three temporal subsets for analysis and evaluation (Figure \ref{fig:graph}) : train set ($G_{train}$), validation set ($G_{val}$), and test set ($G_{test}$). $G_{train}$ contains data from earlier time intervals and is used to expose the model to historical patterns and dynamics within the network. This phase is crucial for the model to analyse historical trends in fish movements. $G_{val}$, consisting of data from a more recent timestamp than $G_{train}$ but preceding $G_{test}$, helps fine-tune the model's parameters to prevent overfitting and improve accuracy.\\
After dividing the data into temporal subsets, similarity scores between node pairs were calculated using various Katz indices to estimate the likelihood of link formation. The similarity scores from the Katz indices were calculated and normalised to facilitate the comparison between the different models. Additionally, the study combined scores from different Katz indices and evaluated their performances.

\subsection{Data}
The data for this study was provided by the Centre for Environment, Fisheries and Aquaculture Science (Cefas)’s Fish Health Inspectorate (FHI). The data included an extensive overview of live fish movement records from fish farms across England and Wales from 2010 to 2023. This dataset captures a total of 16,946 live fish movements, forming a network of 2,480 nodes connected by 4,696 links. Each record in the dataset is associated with the year of the fish movement, allowing for analysis of the network’s evolution over time. Furthermore, the dataset provides detailed information about each movement, including the source and destination farm identifications, geographical coordinates (longitude and latitude), and the farmed species. The dataset was divided into three segments for the model development and evaluation: a training set (2010–2021), a validation set (2022), and a test set (2023). The training set consisted of 2,477 nodes and 4,437 edges. The validation set included 621 nodes and 853 edges, with 3 nodes unique to this set and not present in the training set. The test set, which was used for the final evaluation, comprised 496 nodes and 677 edges.\\
To determine the spatial relationships between farms, we calculated the haversine distance, which takes into account the spherical shape of the earth and is therefore suitable for measuring distances between farms. The haversine distance was chosen because fish farms are often located in various geographical areas, including coastal and inland regions where direct road distance data may not be easily accessible or relevant. While haversine distances provide a more general approximation compared to precise road distances, they are appropriate for this study, due to the considerable distances between farms that require consideration of the Earth’s curvature. 

\subsection{Katz Index}
The Katz index (KI) \cite{katz1953new} is a walk-dependent index, focusing on both direct and indirect walks between nodes. KI directly sums over the collection of unweighted walks between two nodes in a network and is exponentially damped to give shorter walks more weights. The KI is expressed as:

\begin{flalign}
    & KI_{(u,v)} = \sum_{l=1}^{\infty} \beta^l |\text{walks}_{(u,v)}^{\langle l \rangle}| = \sum_{l=1}^{\infty} \beta^l (A^l)_{(u,v)} = \beta A_{(u,v)} + \beta^2 (A^2)_{(u,v)} + \beta^3 (A^3)_{(u,v)} + ..., \label{eq:1} &
\end{flalign}
where $|walks_{(u,v)}^{\langle l \rangle}|$  is the set of length, $l$ walks between node u and v, $A^l$ is the $l^{th}$ power of the adjacency matrix of the network, $\beta$ is the damping factor where a small $\beta$ value means that longer walks contribute less to the Katz index score and vice versa. The choice of $\beta$ must be lower than the reciprocal of the largest eigenvalue of matrix, $A$, to ensure convergence.

\subsection{Modification of the Katz Index}
\subsubsection{Weighted Katz Index}
The weighted Katz index (WKI) modifies the original Katz formula by introducing a weighting factor. This weight, the distance between fish farms in this study, allows the model to account for the varying degrees of distance interaction or influence in the network. The weights are then summed, with a decay factor applied to longer walks to ensure that shorter walks retain higher significance in the prediction model. The model reads:

\begin{flalign}
& WKI_{(u,v)} = \sum_{l=1}^{\infty} \beta^l (A^l_{w})_{(u,v)} = \beta (A_w)_{(u,v)} + \beta^2 (A^2_w)_{(u,v)} + \beta^3 (A^3_w)_{(u,v)} + ...,  \label{eq:3} &
\end{flalign}
where $A_w$ is the weighted adjacency matrix of the network which details the distance between farms.

\subsubsection{Edge weighted Katz Index}
In this study, we developed the edge weighted Katz Index (EWKI), a novel modification of the traditional Katz Index for link prediction in spatially embedded networks.  This modification integrates the distance information between farms as weight, $\omega$. This weight is an exponential function of the distance, $d_{(u,v)}$, between nodes $u$ and $v$, modulated by a decay factor $\gamma$. The inclusion of this weight ensures that the significance of a potential link between two nodes is modulated based on their geographical proximity. As the distance between nodes increases, the $\omega$ decays exponentially, reflecting the reduced likelihood of a link existing between distant nodes. The model reads:

\begin{flalign}
& EWKI_{(u,v)} = \omega_{(u,v)} \cdot \sum_{l=1}^{\infty} \beta^l |\text{walks}_{(u,v)}^{\langle l \rangle}| \label{eq:6} &
\end{flalign}

\begin{flalign}
& \omega_{(u,v)} = e^{-\gamma \times d_{(u,v)}} \label{eq:7} &
\end{flalign}
where $\omega_{(u,v)}$ is the exponential decay weight, $d_{(u,v)}$ is the distance between nodes $u$ and $v$, and $\gamma$ is a constant for the fine-tuning of how quickly the weight decreases with distance, making the model adaptable to various types of networks.

\subsection{Evaluation metrics}
Choosing the right evaluation metrics is essential for comparing the performance of the models, particularly in scenarios like ours where the dataset exhibits sparsity and class imbalance between existing and non-existing links. This study uses precision, recall, F1-score, area under the receiver operating characteristic curve (AUROC) and area under the precision-recall curve (AUPR) value to evaluate the performance of the model.

\begin{enumerate}
    \item Precision: measures the fraction of correctly predicted positive links (i.e., actual fish movements) to all predicted positive links. In the context of our study, a high precision indicates that the model's predictions are predominantly accurate.
        \begin{flalign}
            & \text{Precision} = \frac{TP}{TP + FP} \label{eq:8} &
        \end{flalign}
        where TP = true positive, FP = false positive

    \item Recall: measures the fraction of actual positive links that the model correctly identifies. A high recall in our study suggests that the model captures most genuine fish movements.
        \begin{flalign}
            & \text{Recall} = \frac{TP}{TP + FN} \label{eq:9} &
        \end{flalign}
        where TP = true positive, FN = false negative

    \item F1-score: score is the harmonic mean of precision and recall, which means it gives a balanced measure of the two metrics. An F1-score close to 1 indicates both good recall and good precision, while an F1-score close to 0 indicates poor performance on both metrics.
        \begin{flalign}
            & \text{F1-score} = 2 \times \frac{\text{Precision} \times \text{Recall}}{\text{Precision} + \text{Recall}} \label{eq:10} &
        \end{flalign}

    \item AUROC: is a graphical representation of the trade-off between the TP rate (recall) and the FP rate at various threshold settings. The AUC value, which is the area under this curve, provides a single measure to summarise the ROC curve. It is valuable for comparing the overall performance of different models, irrespective of the threshold chosen.

    \item AUPR value, derived from the AUPR curve, is informative in link prediction where there is a significant imbalance between the positive and negative classes. AUPR focuses more on the positive class, making it a more relevant measure in such imbalanced contexts.
    
\end{enumerate}

\subsubsection{Optimal threshold}
In the evaluation of models, particularly in the context of link prediction, the process of converting calculated scores into binary classification decisions is important. This is usually done through thresholding and is challenging due to the high class imbalance often present in such networks \cite{clauset2008hierarchical, lichtenwalter2010new, wang2007local}. In cases where predetermined classification thresholds are unavailable, it is common to use threshold curves like ROC (Receiver Operating Characteristic) and PR (Precision-Recall) to determine the optimal threshold and assess model performance. However, when dealing with imbalanced datasets, the use of the ROC curve can result in overly optimistic results, since it may not effectively identify rare positive instances effectively. Conversely, the AUPR curve is better suited for imbalanced datasets, as it emphasises the precision-recall trade-off. Here, for each Katz-based score feature, we calculated the F1-scores at various threshold levels. The F1-score, which represents the harmonic mean of precision and recall, offers a balanced metric that is suitable for datasets with significant class imbalances. We identified the threshold that yielded the highest F1-score as the optimal threshold for each feature. By applying these optimal thresholds in our model, we were able to distinguish between predicted links and non-links. This approach not only allowed to establish thresholds that optimise model performance but also facilitated a detailed comparison among different Katz-based models.

\section{Results \& Discussion}
The analysis of various Katz Index-based models on the live fish movement network provided distinct results based on the evaluation metrics from the test data (Table \ref{tab:model_performance} to \ref{tab:confusion_matrix_wkiewki}). The KI model, with an optimal threshold of 0.160, achieved a precision of 0.108 and a recall of 0.208, resulting in an F1-score of 0.142. This model identified 141 TPs, while incurring 1164 FPs and 536 FNs, indicating the model's modest ability to identify true links but struggles with a high rate of FPs (Table \ref{tab:confusion_matrix_ki}), indicating a tendency to over-predict links. The WKI model's performance slightly improves in precision to 0.164 and decreases in recall to 0.123, with an F1-score of 0.140. It only marginally enhances precision while reducing recall, indicating a struggle to capitalise fully on the weighting factor of the model to balance FPs (422) against TPs (83) (Table \ref{tab:confusion_matrix_wki}). EWKI demonstrates a significant increase in precision to 0.988 and recall to 0.712, resulting in an F1-score of 0.827. The model outperforms other models, indicating its high effectiveness in link prediction by identifying most true links (482) with minimal FPs (6) (Table \ref{tab:confusion_matrix_ewki}).\\
For the combined models, KIWKI marginally improves the performance of KI with a precision of 0.128 and recall of 0.276, resulting in an F1-score of 0.175. This combined model aims to leverage the strengths of both KI and WKI and identifies a moderate number of true links (187 TPs), but still suffers from a high number of false positives (1,276 FPs) (Table \ref{tab:confusion_matrix_kiwki}). KIEWKI combines the strengths of KI and EWKI, resulting in a model that effectively balances a high rate of TPs (592) with relatively few FPs (334) (Table \ref{tab:confusion_matrix_kiewki}). The model achieved a precision of 0.639, a recall of 0.874, and an F1-score of 0.739. Lastly, the WKIEWKI, which combines the weighted approaches of WKI and EWKI, achieved a precision score of 0.328 and a nearly perfect recall score of 0.999, resulting in an F1-score of 0.494. This model prioritises recall over precision. Although it identifies almost all true links (676 TPs), it also generates a significant number of false positives (1,386) (Table \ref{tab:confusion_matrix_wkiewki}), indicating a model that emphasises high recall and ensures no true link is missed.

\begin{table}[H]
    \centering
    \caption{Comparative Performance Metrics of Link Prediction Models}
    \renewcommand{\arraystretch}{1.5}
    \begin{tabular}{l c c c c c c} \hline
         Model & KI & WKI & EWKI & KIWKI & KIEWKI & WKIEWKI\\ \hline  
         Threshold & 0.160 & 0.088 & 0.430 & 0.094 & 0.225 & 0.093 \\
         Precision & 0.108 & 0.164 & 0.988 & 0.128 & 0.639 & 0.328 \\
         Recall & 0.208 & 0.123 & 0.712 & 0.276 & 0.874 & 0.999 \\
         F1-Score & 0.142 & 0.140 & 0.827 & 0.175 & 0.739 & 0.494 \\
         AUPR & 0.110 & 0.104 & 0.970 & 0.105 & 0.752 & 0.898 \\
         AUCROC & 0.982 & 0.970 & 1.000 & 0.982 & 0.999 & 1.000 \\ \hline
    \end{tabular}
    \label{tab:model_performance}
\end{table}

\begin{table}[H]
\centering
\caption{Confusion Matrix for KI Model}
\renewcommand{\arraystretch}{1.5}
\begin{tabular}{ccc}
\hline
& Predicted Positive & Predicted Negative \\
\hline
Actual Positive & 141 (TP) & 536 (FN) \\
Actual Negative & 1,164 (FP) & 243,679 (TN) \\
\hline
\end{tabular}
\label{tab:confusion_matrix_ki}
\end{table}

\begin{table}[H]
\centering
\caption{Confusion Matrix for WKI Model}
\renewcommand{\arraystretch}{1.5}
\begin{tabular}{ccc}
\hline
& Predicted Positive & Predicted Negative \\
\hline
Actual Positive & 83 (TP) & 594 (FN) \\
Actual Negative & 422 (FP) & 244,421 (TN) \\
\hline
\end{tabular}
\label{tab:confusion_matrix_wki}
\end{table}

\begin{table}[H]
\centering
\caption{Confusion Matrix for EWKI Model}
\renewcommand{\arraystretch}{1.5}
\begin{tabular}{ccc}
\hline
& Predicted Positive & Predicted Negative \\
\hline
Actual Positive & 482 (TP) & 195 (FN) \\
Actual Negative & 6 (FP) & 244,837 (TN) \\
\hline
\end{tabular}
\label{tab:confusion_matrix_ewki}
\end{table}

\begin{table}[H]
\centering
\caption{Confusion Matrix for KIWKI Model}
\renewcommand{\arraystretch}{1.5}
\begin{tabular}{ccc}
\hline
& Predicted Positive & Predicted Negative \\
\hline
Actual Positive & 187 (TP) & 490 (FN) \\
Actual Negative & 1,276 (FP) & 243,567 (TN) \\
\hline
\end{tabular}
\label{tab:confusion_matrix_kiwki}
\end{table}

\begin{table}[H]
\centering
\caption{Confusion Matrix for KIEWKI Model}
\renewcommand{\arraystretch}{1.5}
\begin{tabular}{ccc}
\hline
& Predicted Positive & Predicted Negative \\
\hline
Actual Positive & 592 (TP) & 85 (FN) \\
Actual Negative & 334 (FP) & 244,509 (TN) \\
\hline
\end{tabular}
\label{tab:confusion_matrix_kiewki}
\end{table}

\begin{table}[H]
\centering
\caption{Confusion Matrix for WKIEWKI Model}
\renewcommand{\arraystretch}{1.5}
\begin{tabular}{ccc}
\hline
& Predicted Positive & Predicted Negative \\
\hline
Actual Positive & 676 (TP) & 1 (FN) \\
Actual Negative & 1,386 (FP) & 243,457 (TN) \\
\hline
\end{tabular}
\label{tab:confusion_matrix_wkiewki}
\end{table}

\subsection{Analysis of AUROC and AUPR Curves}
\begin{figure}[H]
\centering
\begin{subfigure}[b]{0.49\textwidth} 
  \centering
  \includegraphics[width=\linewidth]{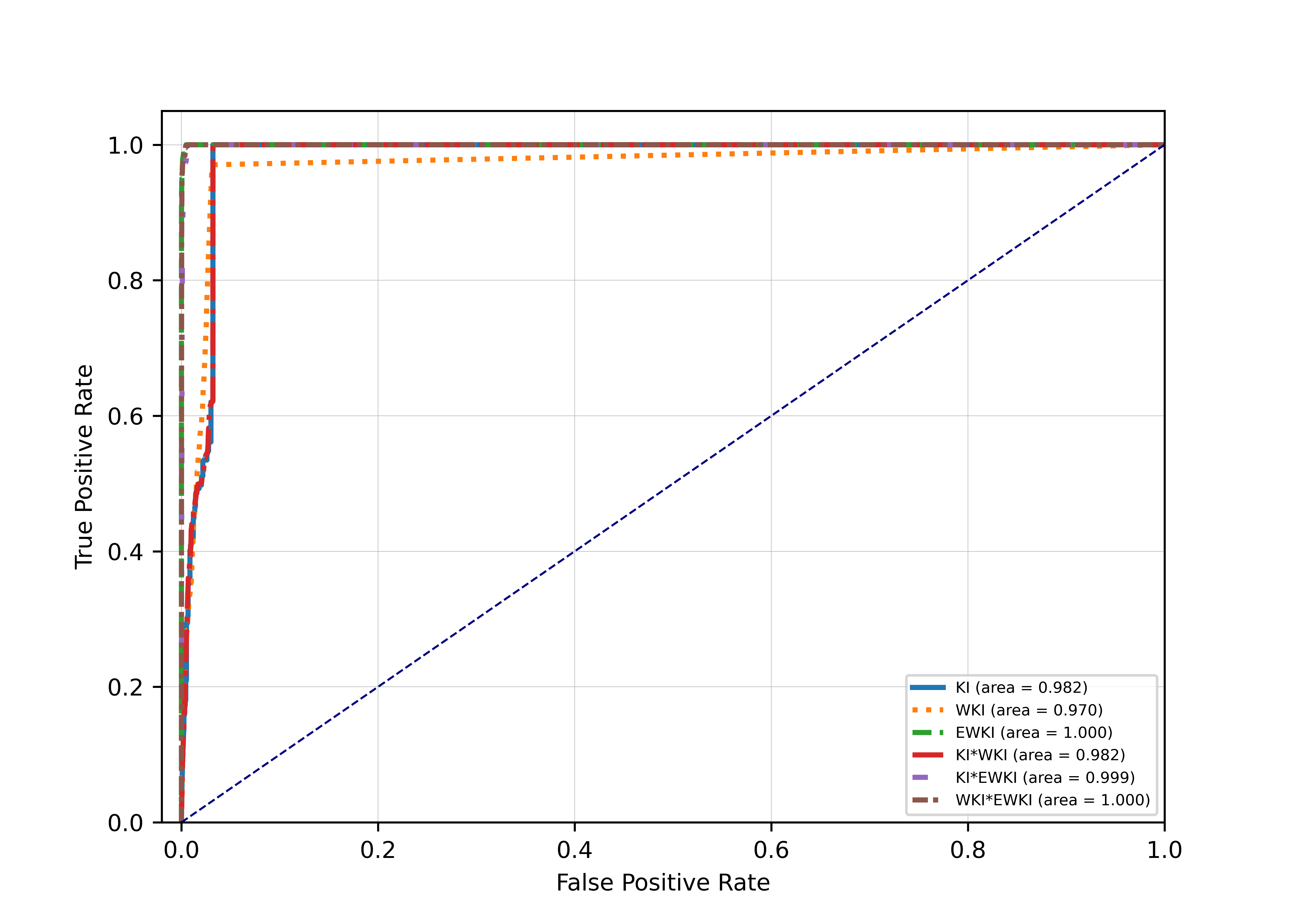}  
  \caption{ROC curves for the various Katz index models.}
  \label{fig:auroc}
\end{subfigure}
\hspace{0.001\textwidth} 
\begin{subfigure}[b]{0.49\textwidth} 
  \centering
  \includegraphics[width=\linewidth]{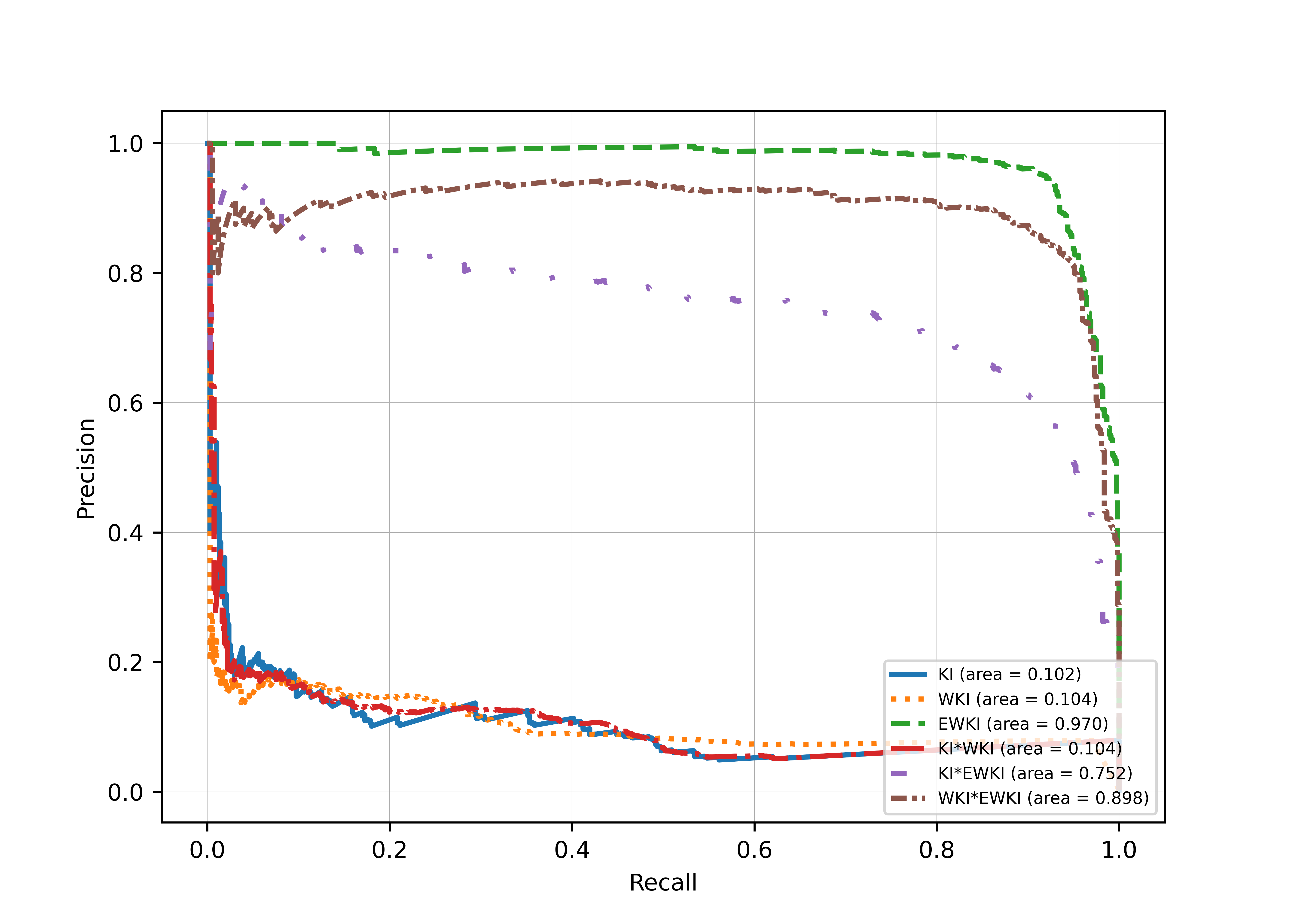}
  \caption{PR curves for the various Katz index models.}
  \label{fig:aupr}
\end{subfigure}
\caption{(a) ROC curves depicting the performance of Katz index models with their respective AUC values, showcasing their ability to distinguish between true and false positives. (b) PR curves representing the precision-recall trade-offs for the same models, highlighting their performance in the context of class imbalance, as measured by AUPR values.}
\label{fig:comparison_plots}
\end{figure}

The AUROC curve measures a model's ability to discriminate between positive (TP) and negative (TP) links across all possible thresholds. A higher AUROC value indicates better model performance, with 1.0 representing perfect discrimination and 0.5 denoting no discriminative ability (equivalent to random guessing). All models exhibited high AUROC values, with EWKI and WKIEWKI achieving perfect scores of 1.000 (Figure \ref{fig:auroc}). This perfection suggests that these models are exceptionally good at ranking positive instances higher than negative ones across all possible thresholds, a crucial capability in link prediction tasks where the goal is to accurately identify potential new links.\\
AUPR is particularly informative in situations of class imbalance, which is a common scenario in link prediction problems. A higher AUPR value indicates that a model not only identifies true positives well (high recall) but also maintains a high level of precision while doing so. The EWKI model's AUPR of 0.970 showcases its efficiency in maintaining high precision across different levels of recall, indicating its robustness in link prediction despite the class imbalance. Conversely, KI, WKI, and KIWKI with lower AUPR values of 0.11, 0.104, and 0.105 respectively (Figure \ref{fig:aupr}), might struggle in environments where precision is as crucial as recall.

\subsection{Implications for Live Fish Movement Prediction}
In aquaculture, disease spread network models can benefit from improved link prediction based on live fish movements. Traditionally, such frameworks have relied on static historical data, which limits their ability to capture the dynamic nature of fish farm interactions. However, using Katz Index-based models can help overcome these limitations. These models take into account the spatial features between fish farms and the temporal changes of live fish movements, resulting in a more accurate representation of disease transmission pathways.\\
Among the evaluated models, the EWKI, KIEWKI, and WKIEWKI models performed exceptionally well in terms of precision, recall, and the AUPR. The EWKI model accurately identified true links, making it an invaluable tool for comprehensive surveillance and early detection of changes in patterns of live fish movement patterns. Its high recall ensures that no potential transmission pathway is overlooked, while its precision minimises the risk of false positives and makes its predictions highly reliable for disease modelling. The KIEWKI model is useful when considering the trade-off between the cost of false positives (unnecessary biosecurity measures and wasted resources) and the risk of false negatives (missed disease transmission pathways and uncontained outbreaks). With its high recall, the KIEWKI model effectively captures nearly all potential transmission pathways, allowing for the early implementation of biosecurity measures. By integrating this model into disease spread models, authorities and farm managers can enhance surveillance and response strategies, adapting them to the dynamic nature of live fish movements and the evolving risk landscape. The WKIEWKI model's near perfect recall rate is valuable in scenarios where missing any link could result in unmonitored disease spread. Its exhaustive detection capability is crucial for developing robust disease models that aim to prevent outbreaks by identifying and monitoring all potential transmission routes. However, the model's moderate precision and higher rate of false positives mean that it may predict connections that do not actually exist.\\
These findings have implications beyond aquaculture and can be applied to other network systems where spatial and temporal interactions are significant. In aquaculture, these models can inform not only disease management, but also logistical and economic decisions related to the movement of live fish. For instance, understanding the likely interactions between farms can provide advice regarding transport organisation, biosecurity measures, and the optimal location for new farm developments. Future work will focus on enhancing these models by incorporating more farm features and advanced approaches, such as graph neural networks, to improve efficiency and prediction accuracy. Additionally, the study aims to explore network rewiring in response to the removal of nodes, which may occur during an outbreak.

\section{Conclusion}
This study aimed to improve the predictive capabilities of link prediction models, specifically focusing on the Katz Index and its variants, based on live fish movements. This approach is key to enhancing the dynamic modelling of disease spread in aquaculture.  Our results have shown that, by incorporating spatial factors, EWKI, KIEWKI, and WKIEWKI perform better than the traditional KI. Geographic proximity significantly influences the movement of live fish as well as disease transmission.

\section{Acknowledgement}
This work was funded by the Department for Environment, Food and Rural Affairs (Defra) [Project FC1215].

\bibliography{Manuscript}

\end{document}